\documentclass[usenatbib]{mn2e}

\newif\ifpdf
	\ifx\pdfoutput\undefined
		\pdffalse % we are not running PDFLaTeX
	\else
		\pdfoutput=1 % we are running PDFLaTeX
		\pdftrue
	\fi

\ifpdf %import graphicx package with options
	\usepackage[pdftex]{graphicx}
	\usepackage[pdftex=true,pagebackref=true,pdfstartview={FitH},pdfpagemode={None},pdfauthor={Graeme Lufkin},pdftitle={Simulations of gaseous disc-embedded planet interaction}]{hyperref}
\else
	\usepackage[dvips]{graphicx}
	\usepackage{hyperref}
\fi

\graphicspath{{figures/}}

\title{Simulations of gaseous disc-embedded planet interaction}

\author[G.~Lufkin et al.]
	{Graeme~Lufkin,$^1$\thanks{Email: {\tt gwl@u.washington.edu}}\thanks{High-resolution figures and movies can be obtained at \url{hpcc.astro.washington.edu/grads/gwl/disc_planet_interaction}}
	Thomas~Quinn,$^1$
	James~Wadsley,$^2$
	Joachim~Stadel$^3$ \newauthor and
	Fabio~Governato$^1$ \\
	$^1$Department of Astronomy, University of Washington, Seattle, WA 98195, USA\\
	$^2$Department of Physics and Astronomy, McMaster University, Hamilton, Ontario L88 4M1, Canada\\
	$^3$Institute for Theoretical Physics, University of Z\"{u}rich, Winterthurerstrasse 190, CH-8057 Z\"{u}rich, Switzerland}

\date{Accepted 2003 September 12.  Received 2003 September 12; in original form 2003 May 26}

\volume{347}
\pagerange{421--429}
\pubyear{2004}

\begin{document}

\maketitle

%The abstract
\begin{abstract}
We present global three-dimensional self-gravitating smoothed particle hydrodynamics (SPH) simulations of an isothermal gaseous disc interacting with an embedded planet.
Discs of varying stability are simulated with planets ranging from 10 Earth masses to 2 Jupiter masses.
The SPH technique provides the large dynamic range needed to accurately capture the large-scale behaviour of the disc and the small-scale interaction of the planet with surrounding material.
Most runs used $10^5$ gas particles, giving us the spatial resolution required to observe the formation of planets.
We find four regions in parameter space: low-mass planets undergo Type I migration; higher-mass planets can form a gap; the gravitational instability mode of planet formation in marginally stable discs can be triggered by embedded planets; discs that are completely unstable can fragment to form many planets.
The disc stability is the most important factor in determining which interaction a system will exhibit.
For the stable disc cases, our migration and accretion time-scales are shorter and scale differently from previously suggested.
\end{abstract}

\section{Introduction}\label{sec:intro}
Planet formation is believed to occur in discs of gas and dust around young stars.
Observations of circumstellar discs in multiple wavelengths reveal the initial conditions for planet formation \citep*{mccaughrean00,wilner00}.
Detection of extrasolar planets has provided a diverse set of final states, including a statistically significant number of planets with masses comparable to that of Jupiter on orbits very close to their parent star, so-called `Hot Jupiters' \citep[for a review see][and {\tt exoplanets.org} for an up-to-date catalogue]{marcy00}.
Formulating an acceptable theory of giant planet formation in such an environment has proved difficult \citep*[and references therein]{wuchterl00}.
Instead planet migration has become the standard theory to explain these interesting companions.

The interaction between a gaseous disc and an embedded protoplanet is a complex, highly non-linear process.
Several phenomena have been classified, including the accretion of planetesimals, the accretion of gas, the exchange of angular momentum via tidal forces and planet formation via gravitational instability.
How these behaviours combine is determined by the initial conditions, for example the mass and temperature of the disc and the location of an embedded planet.
We can sort all possible interactions into a few categories: linear migration via tidal interaction, gap formation and viscous migration, and planet formation via gravitational instability.
All these interactions have previously been investigated separately.
Here we examine how the interactions are related and how the initial conditions determine which behaviour will be exhibited.

The dynamics of low-mass planets in gaseous discs have been explored both theoretically and numerically.
The exchange of orbital angular momentum between disc and embedded satellite via spiral density waves excited at Lindblad resonances was discussed in \citet{goldreich80}.
\citet{ward97} gave arguments suggesting that this will almost always lead to inward migration of the embedded planet, and derived an expression for the rate of this migration in the case where a planet does not open a gap in the disc (Type I migration).
More recently, \citet*{tanaka02} gave an updated formula for the total torque on a planet, taking into account corotation resonances and mentioning three-dimensional (3D) effects.
Previous numerical simulations have tested some of the theoretical assumptions, usually finding general agreement \citep[e.g.][]{nelson00,dangelo02}.

In Type II migration the planet exerts a tidal torque greater than the disc viscous force, opening a gap.
The migration time-scale is slowed, and accretion on to the planet is expected to slow or stop.
Criteria for gap opening have been derived \citep*{takeuchi96,rafikov02} and examined numerically using two-dimensional (2D) grids \citep{bryden99,dangelo02}.
The accretion process has been simulated \citep*{kley99,lubow99} finding mass-doubling times for Jupiter-mass planets in light discs to be of the order of $10^5$~yr.
For a review of Type I and II migration, see \citet{thommes02}.

The standard picture of migration is situated in a calm, stable disc.
What if a single planet forms in a marginally stable disc?
First hinted at by \citet{armitage99}, the spiral density waves excited by a compact object (e.g. a rocky core formed via accretion of planetesimals) can lead to clump formation that would not otherwise occur.
The fact that an embedded planet affects the disc density and flow locally has been known for some time; here we find that it can also have far-reaching (spatially and temporally) global effects.

Regardless of the embedded planets, some discs are themselves gravitationally unstable.
This scenario of planet formation has been explored by \citet{boss97,boss01} and recently by \citet{mayer02}.
In unstable discs many planets of Jupiter-mass or greater can be formed very quickly (hundreds of years).
The newly formed proto-giant planets will violently tear through the disc, and their interaction as point masses will dominate the subsequent evolution of the system.
This scenario is attractive because it easily produces planets with properties similar to observed extrasolar planets.
Furthermore, because the time-scale of formation is so small, any system even marginally unstable is likely to form planets during the much longer disc lifetime.

Much of the previous numerical work in this area has been limited to two-dimensional approximations, using grids with a relatively small spatial range, and ignoring the self-gravity of the disc.
Some three-dimensional simulations have been performed recently \citep{bate03,dangelo03}, although these have still ignored self-gravity.
The smoothed particle hydrodynamics (SPH) formalism was designed to be spatially adaptive.
Our implementation uses variable time-steps, giving us temporal adaptivity.
This allows us to explore the interaction with a much greater dynamic range.
Our simulations are fully three-dimensional, have free boundary conditions, include the self-gravity of the disc and allow the planet to migrate and accrete freely.

Armed with these useful tools, we have simulated planets embedded in gaseous discs, for initial planet masses ranging from 10 Earth-masses to 2 Jupiter-masses, and disc masses ranging from 0.01 to 0.25 $\mathrm{M}_\odot$.
Previous investigations have focused on just one of the interactions described above, usually looking at lower-mass discs.
Here we attempt to relate the interactions, and see how the choice of initial conditions determines the classification of the outcome.
We draw an outline of interesting regions of parameter space, and present a few early results from each distinguishable region.
We also discuss where our simulations differ from previous efforts and suggest that this is due to the more realistic system we are able to model.
In future papers we will quantitatively address the specific phenomena we can simulate, such as gap opening criteria, gas accretion through the gap, Type I and II migration time-scales and the likelihood of triggered planet formation.

The plan of the paper is as follows.
In Section~\ref{sec:initial} we describe the initial conditions of our models.
The numerical technique is briefly discussed in Section~\ref{sec:numerical} and our main findings are detailed in Section~\ref{sec:results}.
Details of the results are discussed in Section~\ref{sec:discussion} and we conclude with Section~\ref{sec:conclusion}.

\section{Initial Conditions}\label{sec:initial}
We use power-law distributions for density and temperature profiles in our simulations, and the disc stability criterion and mass ratio are dimensionless, so our results should scale well.
None the less, it is convenient to use conventional units when discussing our results.
The central star was of 1~$\mathrm{M}_\odot$, and its potential was softened with a cubic spline on a length-scale of 0.4~AU.
All simulations were performed in heliocentric coordinates, taking into account the back-reaction of the disc and planet on the star.
Initially, the disc extends from 1 to 25~AU and is free to expand.

\subsubsection*{\label{sec:temperature_profile} Temperature profile}
Observations of protoplanetary discs do not put tight constraints on the temperature profile.
Therefore, we adopt a power law for the radial temperature profile and choose an index which is equivalent to a constant aspect ratio $H/r$ (where $H$ is the scaleheight of the disc at a particular radius).
Most of our simulations used $H/r~=~0.05$ (which gives $T(5.2~\mathrm{AU})~=~102$~K).
This profile is similar to that suggested by the solar nebula model of \citet*{hayashi85} and is common in other simulations of these phenomena.
The temperature is capped at 5000~K within 0.1~AU of the star and limited below by 17~K outside of 30~AU.

\subsubsection*{\label{sec:density_profile} Density profile}
The minimum-mass solar nebula model suggests a power law for the vertically averaged surface density of $\Sigma(r) \propto r^{-3/2}$.
The vertical structure of the disc is determined by the balance between the vertical component of the gravity of the central star and the vertical pressure support of the disc.
If we assume no vertical variation in the temperature, the vertical density structure can be solved, yielding a scaleheight expressed as a function of the radius and the temperature at that radius.
An expression for the density distribution is then
\[
\rho(r, \theta, z) = \Sigma(r) \rho_{\mathrm{vert}}(r, z) = \left( \Sigma_0 r^{-3/2} \right) \left( \rho_0 e^{-z^2 / H(r)^2} \right)
\]
In a gravitationally stable disc we found this profile to be stable over many dynamical times, so we chose this initial distribution for all of our simulations.
Our standard disc, 0.1~$\mathrm{M}_\odot$ extending from 1 to 25~AU, thus has $\Sigma(5.2~\mathrm{AU})~=~150$~g~cm$^{-2}$.
For the gas particles of the disc, the initial positions are randomly drawn from the desired distribution.
%Our code determines particle timesteps based on the local density, so particles initially placed too close together (Poisson noise) will slowly push their neighbor away with SPH forces.
%This prevents the need for rigid initial conditions such as concentric rings of evenly spaced particles.
Initial velocities are chosen so that particles are on circular Keplerian orbits, modified slightly to account for the self-gravity of the disc and gas forces.
In stable discs without embedded planets, the surface density does not change markedly with time except at the inner and outer edges.
The outer edge expands slightly, due to the initial sharp drop in density.
At the inner edge, particles migrate inward due to viscous forces, participating in a complex balance between the softened potential of the star, the smoothed gas pressure forces and Keplerian shear.
These particles are much closer to the central star than the inner Lindblad resonance of the planet, so we do not believe they play a large role in determining the evolution of the planet.
We have not investigated schemes for halting planet migration, which might involve removing the inner reaches of the disc.

\subsubsection*{\label{sec:planet} The embedded planet}
The initial mass of the planet particle was a parameter.
We used planet masses ranging from 10~$\mathrm{M}_\oplus$ to 2~$\mathrm{M}_\mathrm{J}$.
The planet particle interacts via gravity only, and no gas drag is applied.
Planets significantly more massive than a gas particle can capture gas particles, adding to the effective mass of the planet.
These captured gas particles are not removed or treated specially; they remain as part of the envelope of the developing planet.
The potential of the planet was spline-softened on a length-scale of one-fifth the initial Roche radius.

\subsection{\label{sec:gap} An initial gap}
Sufficiently large planets can open a gap in the disc by exerting a tidal torque stronger than the local viscous angular momentum transfer.
The balance between these forces determines the shape and width of the gap formed around such a planet, which can be calculated analytically by averaging over many orbits and assuming how and where the density waves are dampened.
The formation of these gaps has been studied numerically before, primarily in light discs with heavy planets without allowing migration \citep{bryden99}, finding reasonable agreement with analytic models.
For low-mass discs, the Type I migration time-scale may be sufficiently long for the standard explanation of gap formation to work.
We found that planet migration and gap formation could not be decoupled.
Therefore, we chose not to impose any initial gap, instead the formation of the gap is simulated in systems that produce one.

A Jupiter-mass planet embedded in an unperturbed disc without a gap is somewhat unphysical: how did the planet become so massive without affecting the disc?
As seen in Fig.~\ref{fig:gap_migration} below, the transition from Type I to Type II migration is very sudden, taking only a few orbital times.
Therefore, the sudden appearance of a large planet in a calm disc is not entirely unreasonable, and we can imagine the following scenario: the planet forms, at a lower mass, further out in the disc, and undergoes Type I migration and accretion to obtain the location and mass we start with.
To accurately model the capture of gas, the planet particle must be several times more massive than the gas particles.
Therefore, future investigations using higher resolution should be able to push the initial planet mass lower.

\subsection{\label{sec:eos} Equation of state}
We use an isothermal equation of state with a spatially fixed temperature profile.
In doing so we are assuming that the disc temperature is controlled solely by the distance from the parent star, and that the energy of viscous heating is radiated away much faster than the dynamical time-scale of the gas.
We do not account for shielding, so we miss the detailed thermal structure of the clumps that form.
Furthermore, clumps on elliptical orbits will heat up and cool down along their orbits.
\citet{boss02} has found that using a more realistic equation of state suppresses clump formation in marginally stable discs.

\subsection{\label{sec:Q} Disc stability}
The stability of the disc is characterized by the Toomre $Q$ criterion, defined as
\[
Q = \frac{c_s \kappa}{\pi G \Sigma}
\]
where $c_s$ is the sound speed, $\kappa$ is the epicyclic frequency, and $\Sigma$ is the surface density of the disc.
This criterion measures the susceptibility to a local instability.
Global spiral waves can cause the local density to increase, leading to a local collapse.
The power laws we have chosen for the temperature and density make $Q \propto r^{-1/2}$.
Hence we expect instabilities, if produced, to occur in the outer regions of the disc.

\section{Numerical Realization}\label{sec:numerical}
To model the disc--planet system we used {\em Gasoline}, a tree-based particle code originally developed for cosmological simulations \citep{stadel_phd,wadsley03}.
It implements smoothed particle hydrodynamics, a Lagrangian treatment of hydrodynamics for gas.
It has been used previously on solar system-scale problems by \citet{richardson00} and \citet{mayer02}.
The code was modified to apply a fixed temperature profile to the particles.
The equations of motion include the gravity and hydrodynamics of the gas disc.
We do not include effects due to magnetic fields.

\subsection{\label{sec:boundary_conditions} Boundary conditions}
SPH is a Lagrangian technique, so we do not need to impose any boundaries.
All the particles are allowed to move where the forces dictate.
Particles are not deleted or merged when they approach the central star or the embedded planet.

\subsection{\label{sec:params} Simulation parameters}
The tree--cell interaction opening angle we used was $\theta~=~0.7$.
Using a multi-stepping technique, the particle timesteps are picked according to the local density (with $\eta~=~0.1$), and to satisfy the Courant condition (with $\eta_c~=~0.4$).
The largest possible timestep was set to be 2.5~yr, which is 1/50 the orbital time at the outer edge of the disc.
The mean molecular weight of the gas is 2~AMU, corresponding to molecular hydrogen.

\subsection{\label{sec:viscosity} Viscosity}
The SPH formalism treats viscosity differently from the Shakura \& Sunyaev $\alpha$~prescription popular in finite difference schemes.
No `real' viscosity is added to the equations of motion.
Because so little is known concerning the sources of real viscosity, we choose to err on the side of caution by introducing as little viscosity as possible into the system, in an attempt to model the fluid with as high a Reynolds number as possible.
However, some artificial viscosity is necessary to stabilize the integration of the equations of motion.
{\em Gasoline} includes the standard Monaghan treatment of artificial viscosity \citep{monaghan92}, including a Balsara switch that reduces viscosity in regions of strong shear \citep{balsara95}.
Using the SPH formulation of the equations of motion, we can estimate the kinematic viscosity $\nu~=~\bar{\alpha}~c_s~h~/~8$ \citep{nelson98} associated with the artificial viscosity.
This estimate is reasonable when there is little or no shocking (the presence of shocking adds artificial viscosity locally).
There is strong shear in the disc, so the Balsara switch reduces this by a factor of approximately $5$, yielding a viscous time-scale $\tau~=~r^2~/~\nu~\ga~10^4$~yr everywhere for a run with $10^5$~particles in a stable disc.
Since this is at least 5--10 times longer than our simulations, we believe that artificial viscosity will not seriously affect the dynamics of the disc.
To confirm this, we performed two additional simulations where we halved and doubled the standard value of $\bar{\alpha}~=~1$.
The change in migration and accretion rates was approximately 5~per~cent, which is less than the difference due to the random seed used to determine the initial particle locations.
This indicates that artificial viscosity only affects the simulation as a numerical stabilizer, and does not introduce spurious physical effects.

%The SPH formulation of artificial viscosity and the Shakura \& Sunyaev $\alpha$~prescription often used in grid-based simulations are not directly comparable, because they are solving different problems.
%The $\alpha$~prescription models a real viscosity, whose source may be unknown.
%The artificial viscosity in SPH discussed above is a dissipative term in the equations of motion, to stabilize their integration and capture shocks properly.
%Similar dissipative terms exist in grid-based codes, again to stabilize the integration.
%If we chose, we could add a source of real viscosity to the equations of motion in our SPH formulation, even one that acts just as the $\alpha$ prescription.
%For these simulations, we have chosen not to.

\begin{figure}
\includegraphics[width=8.5cm]{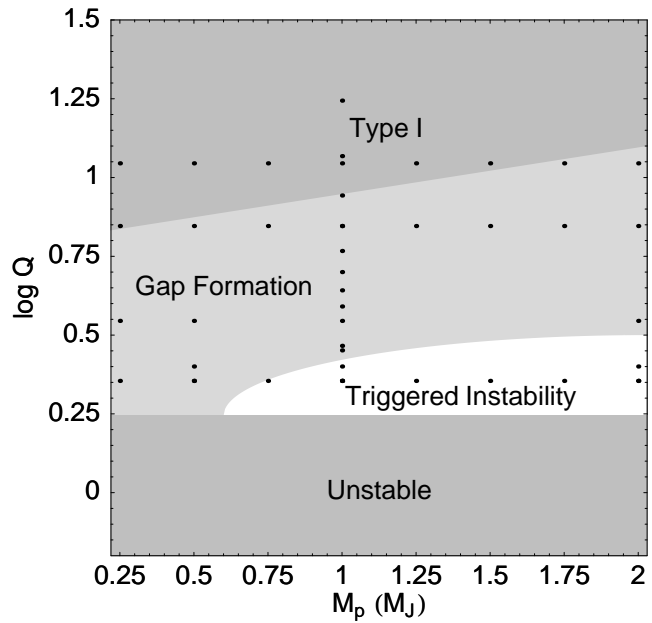}
\caption{A parameter space map of disc--planet interactions.
$Q$ is the initial value of the Toomre criterion at the initial location of the planet.
$M_p$ is the mass of the planet, measured in Jupiter-masses.
The points show the parameter values for the simulations we ran.
See the discussion in the text.
Since the totally unstable discs do not require planets, we cannot technically place them on this plot.
However, if the disc is unstable, an additional planet will certainly not prevent instabilities from growing, so it is reasonable to label a region of this plot as unstable.
}
\label{fig:phase_space}
\end{figure}

\section{Results}\label{sec:results}
Looking first for planet migration, we start with planets embedded in light, stable discs.
Increasing the disc and planet masses increases the rate of Type I migration, and brings us to gap formation and Type II migration.
Continuing to marginally stable discs, we find that the density waves excited by an already formed planet can trigger gravitational instability in discs that would otherwise remain stable.
Finally, we push the disc mass high enough so that no external trigger is needed to drive the gravitational collapse mode of planet formation.

Some of the factors that determine which category a system will end up in include the disc density profile and total mass, disc temperature, initial planet mass and orbital radius.
We made choices for the density and temperature profiles, and varied the total disc mass, the initial planet mass and the orbital radius.
We ran a simulation for each variation of the parameters for a given amount of time (326~yr), and classified the final result.
We found the disc stability to be the most influential factor in determining the behaviour of a disc--planet system.
The qualitative results of our suite of simulations can be seen in Fig.~\ref{fig:phase_space}.
We have arranged our results in a map of parameter space, illustrating the relative positions of the four interactions.

In stable discs (the upper region), planets of all masses exhibit Type I migration.
In more massive discs, planets can open a gap, with a bias toward more massive planets.
A precise measurement of the interface between these two regions can show us the criteria for gap opening.
To differentiate between the regions labelled `Type I' and `Gap Formation' we look for the halting of the migration of the embedded planet within the time we simulated.
See Sec.~\ref{sec:gap_definition} for our definition of a gap and the reasoning behind it.

Some planets can trigger the formation of more planets via gravitational instability.
This requires a marginally stable disc and a massive planet.
The precise shape we have given this region is a guess based on our experience.
In the bottom region, the disc is gravitationally unstable, even in the absence of embedded planets.

Note that the location of the boundaries between regions in this map are dependent on the time period of interest.
We are deliberately simulating for only a few hundred years, to match the time-scale of triggered and wholly unstable planet formation.
As such, many of the systems we classify as Type I do satisfy the standard thermal and viscous criteria for gap formation, and would form deeper depressions in the density profile and eventually shift from Type I to Type II migration if we evolved them several times longer.
In addition, classification over such a short period of time is important, as we measure very high migration and accretion rates.

We have shown a two-dimensional slice through the three-dimensional parameter space we explored.
While convenient and, we believe, largely accurate, this does hide some of the complexity we found.
For example, we looked for triggered instability by placing planets at different orbital radii in different mass discs such that the value of $Q$ was the same at the location of each planet.
Since $Q$ is dimensionless and we used power laws for the density and temperature, we might expect the same result in each system.
However, the system with closer planet and more massive disc did not fragment, while the further planet and less massive disc system did.
This may be due to edge effects or a non-linear dependence on the effectiveness of $Q$ as a predictor of collapse.

\subsection{\label{sec:type_i} Type I migration}
Linear theory suggests that a seed planet in a light gas disc will undergo Type I migration with a rate that scales linearly with the planet mass and disc surface density.
The planet excites density waves in the disc which torque back on the planet, leading to a change in the orbital angular momentum of the planet.
As a result of the power-law index we chose for the density profile, we always see inward migration.
During Type I migration the planet accretes gas from the disc with the help of the spiral density waves it excites.
The internal waves take angular momentum from the disc, and the external waves give angular momentum back to the disc.
On both sides, mass on nearby orbits is shepherded on to the planet.

\begin{table}
\caption{Migration time-scales and accretion rates for Type I systems}
\label{table:type_i}
\begin{tabular}{cccc}
\hline
$M_p$ & $M_d$ ($\mathrm{M}_\odot$) & $-\dot{a}$ (AU/yr) & $\dot{M}$ ($\mathrm{M}_\odot$/yr) \\
\hline
2 $\mathrm{M}_\mathrm{J}$ & 0.01 & $1.2 \times 10^{-3}$ & $5.3 \times 10^{-6}$ \\
1.75 $\mathrm{M}_\mathrm{J}$ & 0.01 & $1.2 \times 10^{-3}$ & $4.7 \times 10^{-6}$ \\
1.5 $\mathrm{M}_\mathrm{J}$ & 0.01 & $1.4 \times 10^{-3}$ & $4.1 \times 10^{-6}$ \\
1.25 $\mathrm{M}_\mathrm{J}$ & 0.01 & $1.1 \times 10^{-3}$ & $3.2 \times 10^{-6}$ \\
1 $\mathrm{M}_\mathrm{J}$ & 0.01 & $1.0 \times 10^{-3}$ & $2.6 \times 10^{-6}$ \\
0.75 $\mathrm{M}_\mathrm{J}$ & 0.01 & $1.4 \times 10^{-3}$ & $1.8 \times 10^{-6}$ \\
0.5 $\mathrm{M}_\mathrm{J}$ & 0.01 & $1.1 \times 10^{-3}$ & $1.2 \times 10^{-6}$ \\
0.25 $\mathrm{M}_\mathrm{J}$ & 0.01 & $8 \times 10^{-4}$ & $2.7 \times 10^{-7}$ \\
1 $\mathrm{M}_\mathrm{J}$ & 0.0035 & $2.7 \times 10^{-4}$ & $6.4 \times 10^{-7}$ \\
1 $\mathrm{M}_\mathrm{J}$ & 0.02 & $2.9 \times 10^{-3}$ & $5.7 \times 10^{-6}$ \\
1 $\mathrm{M}_\mathrm{J}$ & 0.03 & $3.7 \times 10^{-3}$ & $8.5 \times 10^{-6}$ \\
1 $\mathrm{M}_\mathrm{J}$ & 0.04 & $6.0 \times 10^{-3}$ & $1.3 \times 10^{-5}$ \\
1 $\mathrm{M}_\mathrm{J}$ & 0.05 & $7.2 \times 10^{-3}$ & $1.9 \times 10^{-5}$ \\
1 $\mathrm{M}_\mathrm{J}$ & 0.06 & $1.0 \times 10^{-2}$ & $2.3 \times 10^{-5}$ \\
\hline
\end{tabular}

\medskip
$M_p$ is the planet's initial mass, $M_d$ is the total disc mass, $\dot{a}$ is the inward migration rate, $\dot{M}$ is the accretion rate on to the planet.
All these planets started out on circular orbits at 5.2~AU.
The rates given are equilibrium values over at least the last 200~years of the simulation (which typically lasted 326~years).
For the lighter discs the long-term values were reached immediately.
We give values for particular simulations, so can give no meaningful error estimate.
However, we completed several runs at different resolution, and usually found agreement within 10~per~cent.
\end{table}

Table~\ref{table:type_i} lists migration and accretion rates for the systems we classified as Type I migration.
For the lighter discs, the migration rate is a constant function of time, whereas a more massive disc causes the migration rate to decrease as the simulation progresses and a gap is formed.
We find that the accretion rate scales linearly with the planet mass and disc surface density.
Further, the migration rate scales linearly with the disc surface density.
However, changing the planet mass has almost no effect on the migration rate, confirming the findings of \citet{nelson03b}.
If we compare our results with the analytic expression (eq. 70 from \citet{tanaka02}), we find our migration time-scales ($\tau_M = a / |\dot{a}|$) to be approximately 6 times shorter.
Most previous simulations of Type I migration have kept the planet on a fixed circular orbit, and estimated migration time-scales by measuring the torque on the planet from the gaseous disc.
Our migration timescales are actual measurements of the rate of change of the orbit of the planet.

\subsection{\label{sec:gap_formation} Gap formation}
More massive seed planets exert strong enough tidal torques to open a gap in the disc.
By clearing the gas from the location of the Lindblad resonances, the torque exerted on the planet by the disc vanishes.
Previous theoretical work and simulations have suggested that formation of a gap drastically reduces the accretion rate on to the planet.
We have not found this to be true.
Even when the gap has cleared, the planet still perturbs the disc.
These perturbations act to pull mass off the edges of the gap and on to the planet.

\begin{figure}
\includegraphics[width=8.5cm]{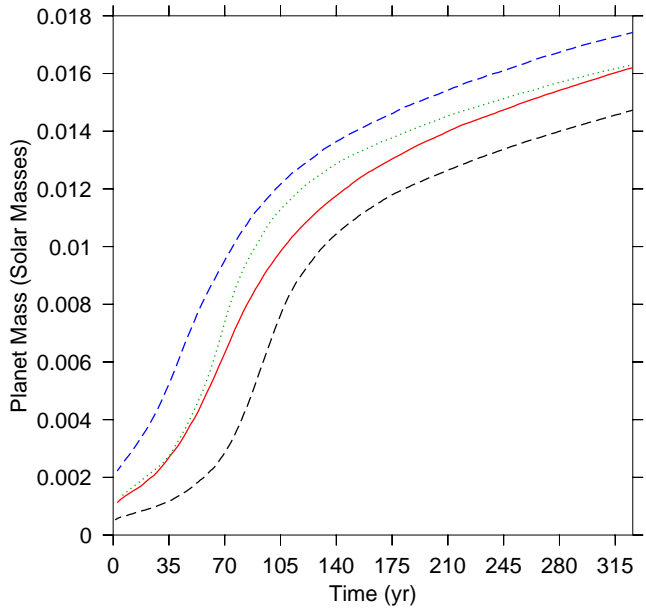}
\caption{Gap formation scenario: The planet mass as a function of time in a 0.1~$\mathrm{M}_\odot$ disc, for different starting masses.
The solid line is for an initial planet mass of 1~$\mathrm{M}_\mathrm{J}$.
The top dashed line is 2~$\mathrm{M}_\mathrm{J}$ and the bottom dashed line is 0.5~$\mathrm{M}_\mathrm{J}$.
The dotted line is for a 1~$\mathrm{M}_\mathrm{J}$ planet in the 2D approximation.
}
\label{fig:gap_accretion}
\end{figure}

\begin{figure}
\includegraphics[width=8.5cm]{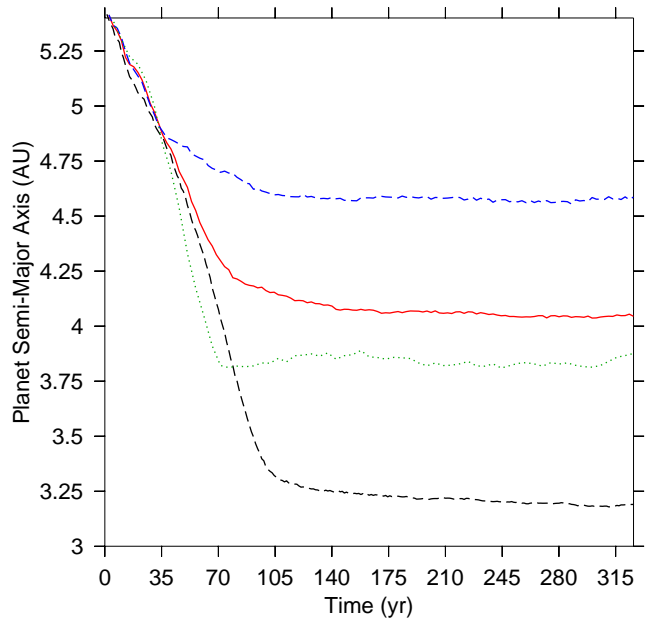}
\caption{Gap formation scenario: The planet orbital radius as a function of time in a 0.1~$\mathrm{M}_\odot$ disc, for different starting masses.
The solid line is for an initial planet mass of 1~$\mathrm{M}_\mathrm{J}$.
The top dashed line is 2~$\mathrm{M}_\mathrm{J}$ and the bottom dashed line is 0.5~$\mathrm{M}_\mathrm{J}$.
The dotted line is for a 1~$\mathrm{M}_\mathrm{J}$ planet in the 2D approximation.
}
\label{fig:gap_migration}
\end{figure}

We found that gap formation was dependent more on the disc mass than the planet mass.
This agrees with our claim that the self-gravity of the disc is important.
At the start of the simulation the planet rapidly collects all the gas particles nearby, forming a gap within approximately 100~years.
This accretion of gas at corotation is much faster than the accretion from close spiral arms or from gap edges, as described above.
Since the migration time-scale is so short in the massive discs, there is significant migration during the gap formation.
After the gap has formed, the migration stops or slows considerably.
In 0.1~$\mathrm{M}_\odot$ discs, the accretion rate slows to a steady value of approximately $10^{-5}$~$\mathrm{M}_\odot$/yr, independent of the mass of the planet.
Fig.~\ref{fig:gap_accretion} shows the mass of a planet and its accreted envelope as a function of time, for 0.5, 1, and 2~Jupiter-mass seed planets.
By comparing with Fig.~\ref{fig:gap_migration}, the semi-major axis of the planet as a function of time, we see the accretion slow as the planet clears its gap (changes from Type I to Type II migration).
The planet peels gas off the edges of the gap, as it catches up with particles on the outer edge and is caught by particles on the inner edge.
Our results confirm that the formation of `Super-Jupiters' is possible in massive discs.
As Type I migration ceases, the planet is left in the gap with a non-zero eccentricity.
We saw eccentricities of 0.01 to 0.04, comparable to the current eccentricity of Jupiter.

The time-scale for Type II migration is the viscous time-scale of the disc.
Since we are actively trying to reduce the viscosity of the disc in our simulation, we do not see significant migration after the gap has formed.

\subsection{\label{sec:triggered} Planet--triggered instability}
\begin{figure*}
\includegraphics[width=6.5in]{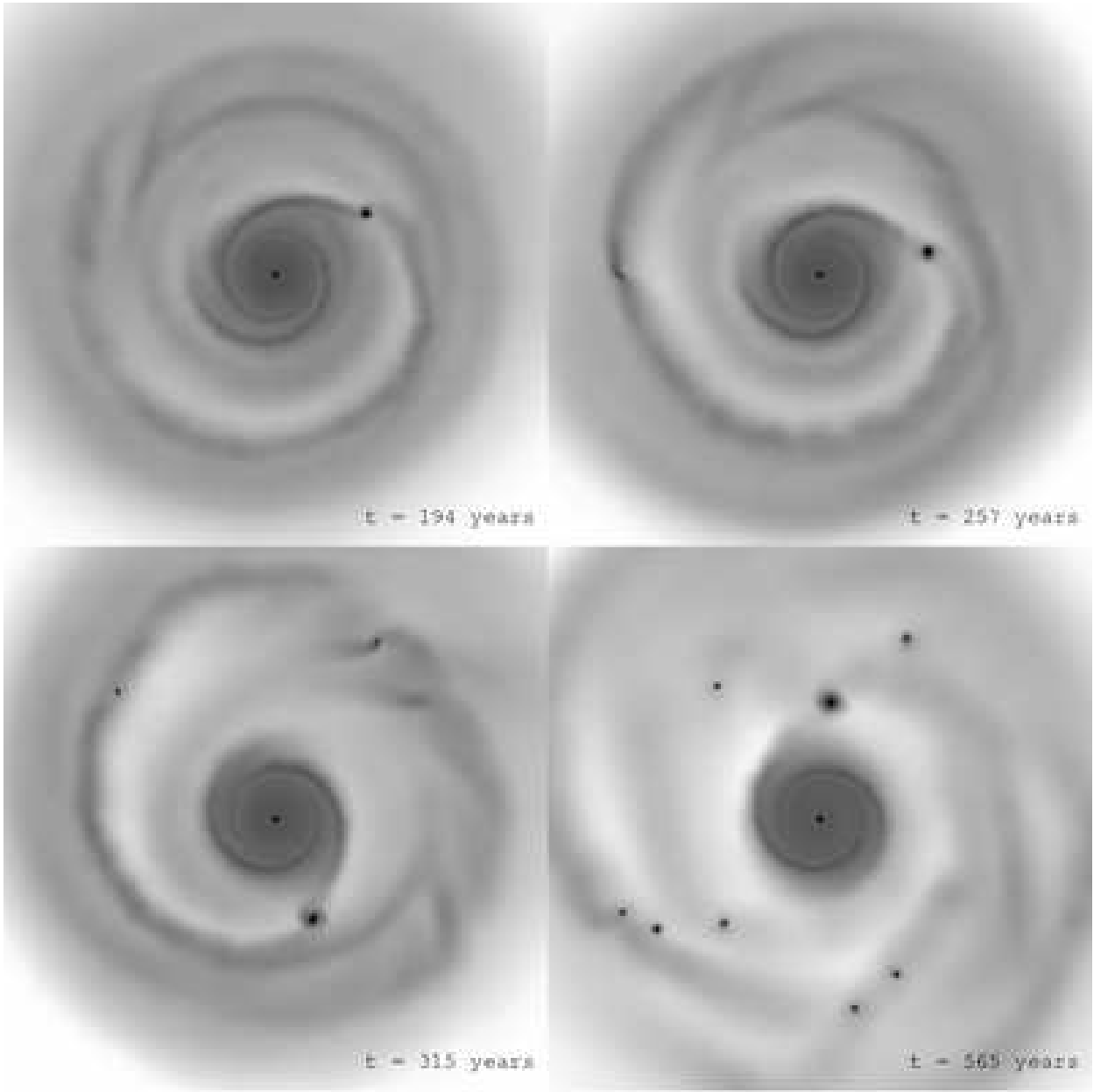}
\caption{{\bf Triggered Planet Formation}: A sequence of snapshots showing the gas density in a protoplanetary disc.
At the planet's location $Q$~=~2.2, at the outer edge $Q$~=~1.6.
Early, the planet drives spiral density waves in the outer regions of the disc.
One of these waves increases the local density to the point where the region collapses.
The planet continues to drive the spiral arms, leading to further clump formation.
The process will continue until the disc material is sufficiently depleted.
The width of each image represents 55~AU.}
\label{fig:triggered_sequence}
\end{figure*}

If we reduce the stability of the disc, the spiral arms generated by an embedded protoplanet can trigger the gravitational instability.
Such triggered clumps are always exterior to the seed planet that drives the instability, because $Q$ decreases with distance from the star.
The spiral arms of the planet must present a large enough density contrast to drive the local value of $Q$ below 1.
Since the unperturbed disc stability decreases with distance from the central star, for a fixed disc stability, a planet farther out from the central star is more likely to trigger the collapse.
If we place a Jupiter-mass planet at 12.5~AU, a disc with $Q \ga 1.6$ will fragment (see Fig.~\ref{fig:triggered_sequence}).
Once the instability is triggered, the evolution of the system is very similar to the totally unstable case.
The clumps form with non-negligible eccentricity, and strongly disturb the disc profile.

\subsection{\label{sec:unstable} Unstable discs}
\begin{figure*}
\includegraphics[width=6.5in]{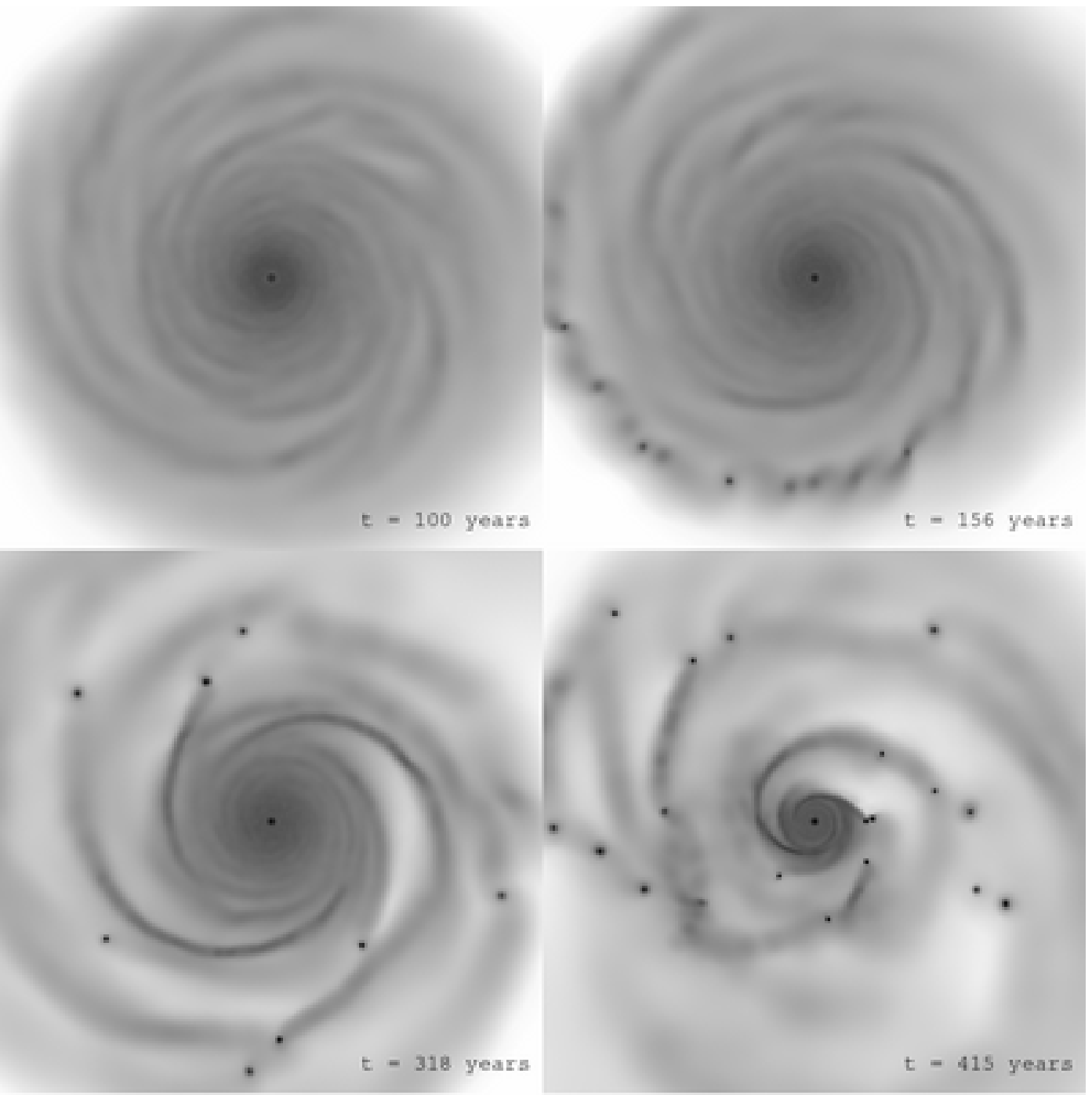}
\caption{{\bf Unstable Disc Planet Formation}: A sequence of snapshots showing the gas density in an unstable disc ($Q_{\min}$~=~0.8).
Early on, the spiral arms grow.
After 150~years one of the spiral arms fragments into a few clumps.
As time goes by, other arms fragment, in the outer disc.
Later, some clumps' leading arms cause fragmentation of the inner disc.
The width of each image represents 55~AU.}
\label{fig:unstable_sequence}
\end{figure*}

If the disc is massive and/or cool enough, gravitational instabilities will cause it to collapse into bound objects, regardless of any embedded planets.
Spiral waves propagate, and the density buildup in the arms causes the disc to fragment into gravitationally bound objects.
Not one, but several of these bound objects are formed, and their interaction as effective point masses dominates the subsequent evolution.
The previously smooth disc is ripped apart, as the clumps often have sizable eccentricities (see Fig.~\ref{fig:unstable_sequence} for a progression of snapshots).
We observe collisions, ejections, tidal stripping during close encounters, and violent disruption when plunging too close to the parent star.
The eccentricities of these clumps can be high, and change as a result of dynamical interaction.
As in \citet{mayer02}, we find that the disc is unstable when Toomre's $Q$ criterion is less than approximately $1.4$.
The clumps form when a spiral arm with sufficient amplitude reaches the outer regions of the disc.

\begin{figure}
\includegraphics[width=8.5cm]{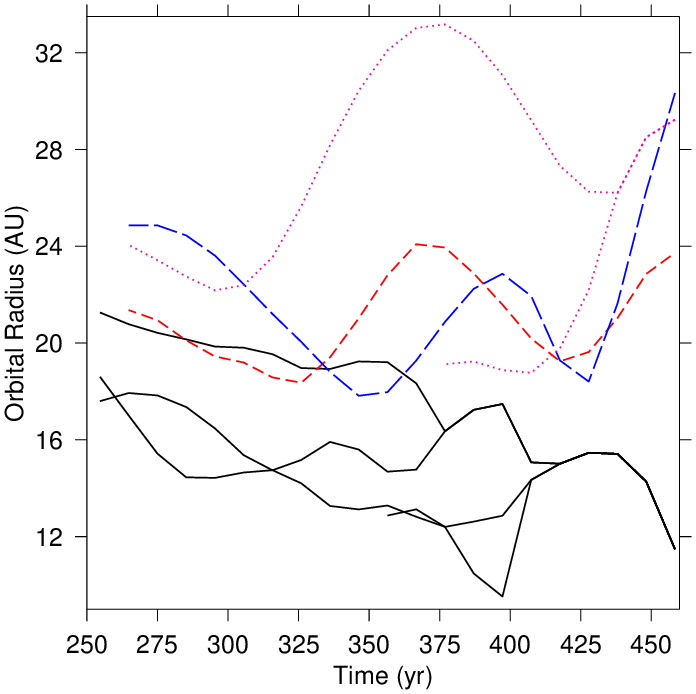}
\caption{Formation in an unstable disc: The orbital radius of several clumps as a function of time.
The lines begin when a clump forms.
Clumps that will merge in the time shown have the same line style.
}
\label{fig:clump_position}
\end{figure}

\begin{figure}
\includegraphics[width=8.5cm]{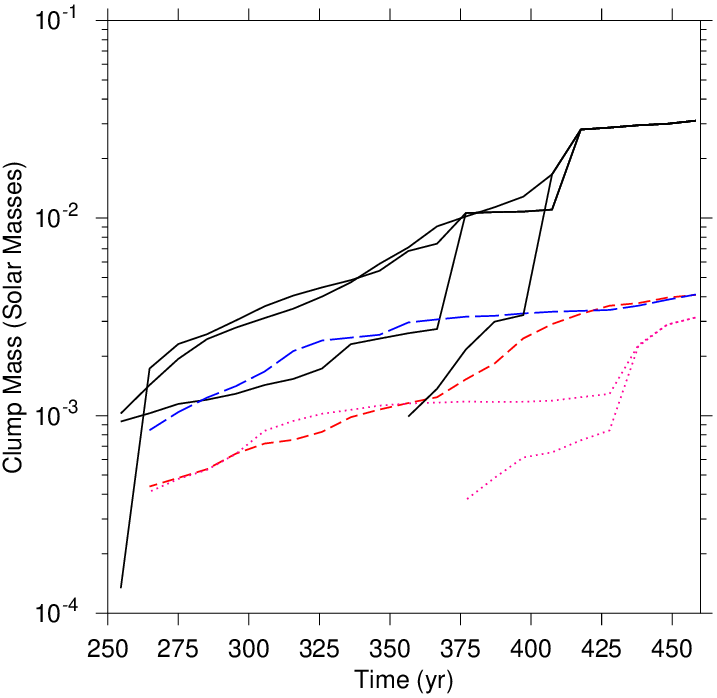}
\caption{Formation in an unstable disc: The mass of several clumps as a function of time.
The lines begin when a clump forms.
Clumps that will merge in the time shown have the same line style.
}
\label{fig:clump_mass}
\end{figure}

For this regime we present a simulation that was unstable, a 0.2~$\mathrm{M}_\odot$ disc which has $Q_{\min} = 0.8$.
In this run, a spiral arm at approximately 20~AU fragments into four separate clumps after 153~years.
Five more clumps form over the next 200~years.
At this point, two clumps have developed strong leading arms, which fragment into approximately 10 more clumps over the next 30~years.
Figs.~\ref{fig:clump_position} and \ref{fig:clump_mass} show the orbital and mass progression of some of the first clumps formed (the simulation forms many more clumps than detailed in the plots).
We see large eccentricities and eccentricity changes, and mergers of several clumps.
Once a significant fraction of the gas has been caught up in clumps, we expect that the furious production of new bound objects will cease, and the system will evolve as a set of point masses.
After a long time (much longer than we have simulated) the system will reach dynamical stability, presumably with only a handful of objects remaining.


When a spiral arm fragments, it collapses to a rotating spheroid.
All our planets formed initially with prograde rotation, however a later merger resulted in a counter-rotating clump.
We are not modeling the cooling of the atmosphere of the planet accurately, so cannot comment on what the final spin of the planets will be.
Because our simulations are three-dimensional, we can and do observe tilted planets, a result of interactions with other planets and vertical asymmetry in the spiral density waves.

Our work on wholly unstable discs extends the work of \citet{mayer02} to a larger region of parameter space, following the system for a longer period of time and spatially fixing the temperature profile.
The critical value of $Q$ that heralds collapse is quite sensitive to the equation of state used \citep{boss02}.
A more detailed equation of state and more physical treatment of radiative transfer are necessary to determine how likely planet formation via gravitational instability is in a given disc.
Note that, though we have evolved the system well past the initial collapse, we do not believe we have accurately modeled the detailed structure of the planets that form because we are using a simple equation of state.
However, we have accurately followed the gravitational interaction of the multiple planets that form.
We can therefore assert that, in planetary systems that formed via gravitational collapse, the disc torque theory of migration does not play a role, as the two-body interactions completely dominate the orbital evolution of the planets.

\section{Discussion}\label{sec:discussion}
Here we describe how we calculated the mass of the planet, some tests we performed, further discussion of our migration results, and how we define and use the word `gap' in this work.

\subsection{\label{sec:planet_mass} Mass of the planet}
The gas particles that are captured by the planet are not removed from the simulation or added to the planet particle.
They continue to orbit in the disc along with the planet.
When we give results for accretion rates we have to calculate a mass of the whole planet, including the initial point mass and the captured gas.
We use a simple criterion to determine whether or not a gas particle counts toward the mass of the planet: a particle must be within the Hill sphere of the planet and have density greater than the Hill density (the planet mass divided by the volume of the Hill sphere).
Since the Hill sphere depends on the planet mass, this process is iterated until the planet no longer grows, starting from the initial planet particle.

\subsection{\label{sec:resolution} Resolution issues}
Our standard run for Type I migration is a 0.01~$\mathrm{M}_\odot$ disc with a Jupiter-mass planet starting on a circular orbit at 5.2~AU.
We simulate the system for 326~years (approximately 28~orbits) and calculate accretion and migration rates as a function of time.
Over this time period the migration and accretion rates are constant, and the planet migrates inward approximately 0.3~AU and grows in mass to approximately 1.8~$\mathrm{M}_\mathrm{J}$.
We ran this model at several resolutions to check for convergence.
At resolutions of 20000 to 400000 particles the accretion rates matched to within 5~per~cent and the migration rates to better than 1~per~cent.
Given this agreement, we chose $10^5$ particles as our standard resolution.

The results of two $10^5$ particle runs can be seen in Fig.~\ref{fig:surface_density}.
Each simulation initially has $\Sigma(r)~\propto~r^{-3/2}$, aspect ratio $H/r~=~0.05$, and an embedded Jupiter-mass planet initially on a circular orbit at 5.2~AU.
The two density profiles shown are for disc masses of 0.1 and 0.01~$\mathrm{M}_\odot$, after 326~years.
The disc is three-dimensional, has free boundary conditions, and self-gravity is included.
The density profiles can be compared to the results of previous simulations, for example fig.~3 of \citet{nelson03b} and fig.~10 of \citet{kley99}.
Those simulations are both two-dimensional, treat viscosity differently, and apply boundary conditions at the inner and outer edges of the disc.

\subsection{\label{sec:explaining_migration} Migration rate differences}
The Type I migration rates we found were higher than most previous numerical and semi-analytical studies.
However they agree fairly well with a recent simulation  of \citet{nelson03b}, which does include the self-gravity of the disc.

We believe our treatment of the vertical dimension and the disc self-gravity is correct, and so the discrepancy should reveal invalid assumptions made in the linear theory of migration.
Self-gravity always acts to increase the magnitude of density perturbations, so we would expect that the torques arising from these perturbations would be correspondingly increased, but we have not completed any simulations without self-gravity, so cannot comment further.
To determine the effect of including the vertical dimension of the disc, we did several runs where all particles were constrained to the $z=0$ plane.
For light discs, the 2D case consistently gave accretion rates lower than the 3D case by at least 20~per~cent.
For more massive discs, the 2D case appears to give a slightly higher accretion rate during gap formation but again a lower rate after the gap has formed (see Fig.~\ref{fig:gap_accretion}).
The difference in migration rates between 2D and 3D simulations was not clear.

\subsection{\label{sec:gap_definition} Defining a gap}
\begin{figure}
\includegraphics[width=8.5cm]{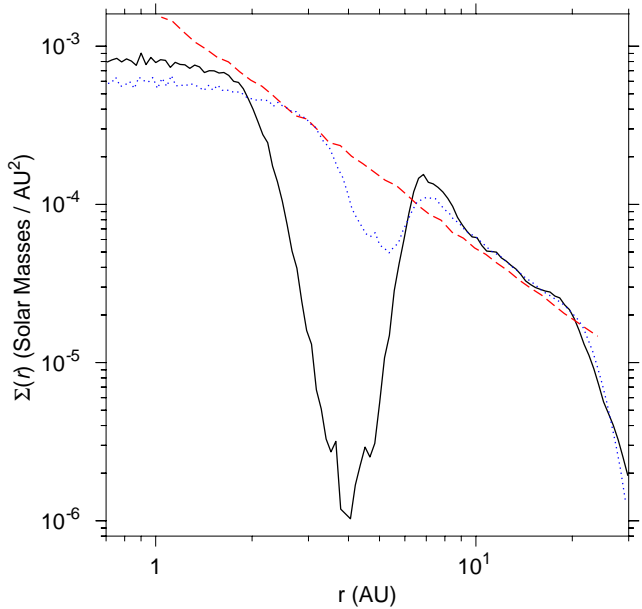}
\caption{Azimuthally and vertically averaged surface density of two discs with embedded Jupiter-mass planets, after 326~years.
The solid line is the density of a 0.1~$\mathrm{M}_\odot$ disc.
A deep gap has formed within 100~years, and Type I migration has stopped.
The dotted line is the density (multiplied by 10) of a 0.01~$\mathrm{M}_\odot$ disc.
Note that a depression has formed about the planet, of less than half the initial density.
This feature can be called a gap, but the planet is still exhibiting Type I migration.
The dashed line shows the initial density profile.
}
\label{fig:surface_density}
\end{figure}

Determining if and when a planet forms a gap in a disc is a non-trivial problem.
By what percentage must the density decrease around a planet to be called a gap, and how far must this density depression reach?
These criteria necessarily invoke an arbitrary number.
Fig.~\ref{fig:surface_density} shows an azimuthally and vertically averaged surface density of two simulations with different disc masses.
In both cases the gas density near the planet has been markedly reduced, and these depressions could be called gaps.
The planet in the heavier disc has stopped migrating (see Fig.~\ref{fig:gap_migration}).
In the lighter disc, however, the planet continues to migrate inward at a rate independent of the depth of the density depression.
Therefore, we choose to use the word `gap' only when its dynamical effects are apparent, i.e. when the planet has stopped migrating.
This is how we chose between the `Type I' and `Gap Formation' regions when classifying simulations for Fig.~\ref{fig:phase_space}.

\section{Conclusion}\label{sec:conclusion}
This work presents the first results from our ongoing investigation of planet formation and migration.
It is a pioneering application of global SPH simulations to a field previously examined primarily with Eulerian grid-based simulations.
Using the new technique, we have measured migration and accretion time-scales, witnessing Type I migration and gap formation.
Our results for this region partially qualitatively agree with the linear theory of interaction with spiral density waves.
We have observed planet formation via disc fragmentation, including a cascade effect from a single starter planet.
Using massive discs, we find that it is easy to form large (tens of Jupiter masses) planets on a short time-scale (hundreds of years).
If planets form via gravitational instability, the theory of migration via disc torques is not applicable.

To obtain an overview of gaseous disc--embedded planet interactions, we performed a suite of simulations, varying the planet mass, the initial location of the planet, and the disc stability.
We have shown the orientation in parameter space of four broad categories of interaction.
These interactions are: Type I migration; gap formation and Type II migration; planet-triggered instability; and unstable discs.
The stability of the disc is the most effective predictor of the type of interaction between a gaseous disc and an embedded planet.

Planet formation and migration is a non-linear process that requires the reasoned application of computational resources to model.
Future investigations will explore the details of each interaction we have described.
In addition, we will refine our knowledge of the boundaries between regions of parameter space.
Future observations will give us a range for $Q$ and a better understanding of the temperature profile in protoplanetary discs.
Combined with the results presented here, we can say which migration and formation processes will be relevant.

%\subsection{\label{sec:future} Future Work}
%With continued application of our simulation technique, we hope to answer many more questions about giant planet formation.
%For unstable discs, we will give tighter constraints on the conditions leading to instability, and attempt to generate a planet mass distribution function, given a range of initial disc configurations.
%Likewise, we will refine our estimates of the conditions necessary for an already-formed planet to trigger the gravitational instability.
%For migration studies, we will examine more closely the distinction between Types I and II migration, determining migration rates for a larger range of planet masses, and check for convergence using higher-resolution simulations.
%
%In choosing our initial conditions, we opted for power laws in the temperature and density profile, so that our results would scale well.
%The particular indices chosen may affect the dynamics, and we intend to check this with more simulations.
%In addition, future observations may give us a better idea of the range of reasonable values.
%
%The equation of state and radiative transfer may prove to be important to the phenomena in which we are interested \citep{boss01,rice03}.
%While we are still far from a full treatment of radiative transfer in our code, we are implementing a system of cooling relevant to solar-system scale problems.

%\section{Acknowledgments}

\bibliographystyle{mn2e}
\bibliography{gwl}

\end{document}